\documentclass[12pt]{iopart}

\usepackage{iopams}  
\usepackage{graphicx}
\renewcommand{\thesection}{\Roman{section}} 
\renewcommand{\thesubsection}{\thesection.\alph{subsection}}
\begin{document}

\title[]{Addressing the anomalies in determining negative ion parameters using electrostatic probes}

\author{Pawandeep Singh$^{1,2}$ and Shantanu Kumar Karkari$^{1,2}$}

\address{$^{1}$Institute for Plasma Research, Bhat, Gandhinagar, Gujarat, 382428 India}
\address{$^{2}$Homi Bhabha National Institute, Training School Complex, Anushakti Nagar, Mumbai-400094, India}

\ead{singh.pawandeep67@gmail.com and  skarkari@ipr.res.in }
\vspace{10pt}

\begin{abstract}
The negative ion density and temperature are the two fundamental parameters that are necessary to quantify the properties of electronegative discharges. However, determining these parameters by means of electrostatic probes can be quite challenging because of the inherent inaccuracies involved in determining the electron/ion saturation currents, electron temperatures and plasma potential, which relies on charge particle collection by the probe surface; as well as on the sheath models that are originally developed for an ideal collision-less plasma. This paper briefly reviews the various limitations associated with these underlying methods and suggests useful means to correct the anomaly associated in determining the negative ion parameters based on electrostatic probes.   
\end{abstract}

\section{Introduction}

Over the past several decades, research on negative ion plasmas has remained in the centre stage due to its overwhelming applications in micro-electronic industries to the generation of energetic neutral particle beams for plasma heating in fusion devices \cite{Franzen2013} to the promising application in plasma propulsion \cite{Aanesland2009}. The research is mainly driven from the need to develop negative ion sources for various applications as well as to study the fundamental properties of bulk plasma and the sheaths associated with its confining boundaries. These plasmas have some unique properties which distinguish it from the more commonly found electron and positive ion plasmas in the laboratories \cite{Franklin2002a}. The reason is because the negative ions are quite massive than its counterpart electrons, hence their contribution to charge shielding inside plasma is more robust than the thermally agile electrons. As a result the ambient electric field inside electronegative plasma is weaker than the plasmas consisting of pure electrons and positive ions. A weaker electric field inside the pre-sheath region means that the positive ion speed/ flux towards the sheath boundary gets sufficiently reduced, whereas the majority population of negative ions tends to remain confined inside the plasma volume. This is one of the reasons to observe low extraction efficiency of negative ions from the bulk plasma. In order to investigate such plasma sources, analytical models supported by experimental data to map the underlying plasma parameters can be quite helpful. Therefore reliable plasma diagnostics is inevitable for the research on negative ion containing plasmas for various applications.

Diagnostics of negative ion parameters is by far the most challenging aspect in negative ion research. Conventionally the electrostatic probe remains the fundamental choice for the experimentalists, as it is simple to construct and it can also provide local measurements. The other prominent methods to diagnose negative ion parameters are based on cavity-ring-down spectroscopy \cite{Bandyopadhyay2019} and pulsed laser-photo-detachment \cite{Conway2010a}, but these techniques are quite cumbersome and expensive. Besides many plasma systems do not have access to introduce the laser beam inside the discharge chamber. In addition, in the case of pulse laser photo-detachment method, the electrical probes are still required to measure the photo-detached electron density during the injection of the pulsed laser beam inside the plasma. Therefore the accuracy of these techniques greatly depends on the performance of the probes introduced inside the plasma for the detection of negative ions.

  The first independent use of Langmuir probes to diagnose negative ions was originally developed by Sheridan et al \cite{Sheridan1999}. This underlying method is based on comparing the ratio of electron to positive ion saturation currents to a cylindrical Langmuir probe to that with pure electron-positive ion plasma. Over the years, several authors have further improvised this concept and tested those in laboratory plasma devices \cite{Bredin2014,Bowes2014,Pandey2017,Sirse2013,Amemiya1999}. It is found that this method is relatively more accurate in highly electronegative discharges. However, in modest to weakly electronegative discharges, the parameters obtained are found to be quite sensitive to the electron /ion saturation currents, the electron temperature, plasma potential, etc. that has to be determined from the Ampere-Voltage (I-V) characteristics of the probe. Furthermore, the plasma parameters obtained from the analysis of saturations current techniques are based on specific probe theories which require a prior knowledge of negative ion parameters as well. In the recent years, alternative methods based on resonance hairpin probe to determine the negative ion density and temperature has also been demonstrated \cite{Sirse2013,Pandey2020,Sirse2015}. However, these techniques are relatively new and also the popularity of the use of hairpin probes has been confined to a limited research groups.

      This work is primarily motivated to address the specific limitations associated with Langmuir probes that are conventionally used for the measurement of negative ion parameters in an electronegative discharge. As highlighted above the accuracy in determining negative ion parameters using Langmuir probe is very much depended on the electron and positive ion saturation currents. For instance Bhuva et al. \cite{Bhuva2019} demonstrated that the electron saturation current gets sufficiently diminished if the reference electrode is poorly in contact with the plasma. The presence of non-thermal electrons can also shift the floating potential to larger negative values relative to the bulk plasma potential. In RF plasmas, determining the plasma potential can be quite tricky due to presence of RF oscillation and presence of external magnetic field. Similarly, it is observed that the positive ion current to a cylindrical probe continues to increase with application of negative probe bias. This effect can introduce significant errors in the estimation of positive ion saturation current as well. Finally the probe tip itself introduces a local perturbation inside plasma, therefore the charged particle distribution around the probe surface needs to be extrapolated to get the negative ion parameters inside the plasma bulk. Furthermore, different authors have interpreted saturation currents from Langmuir probes to quantify negative ions. Therefore the aim of the paper is to present a comparative analysis prescribed by different authors and attempt to address the various anomalies associated with the individual techniques.

In Section-\ref{thr} a brief review on the saturation current method using Langmuir probes in low-pressure electro-negative discharges is presented. The errors while interpreting the sheath edge density around cylindrical probe is discussed in Section-\ref{thrcorr}. Next the errors associated during the measurements of saturation current ratio are addressed in Section-\ref{experr}. Section-\ref{ExpRes} presents the experimental setup and results, the methods adopted by varies authors for the determination of electronegative parameter, $\alpha$ and compares the results with an improved method. Finally the important findings are summarized in Section-\ref{Sum}.

\section{Langmuir probe to determine negative ion density }\label{thr}
\begin{figure*}[tbp]
  \centering
     \includegraphics[width=0.5\textwidth]{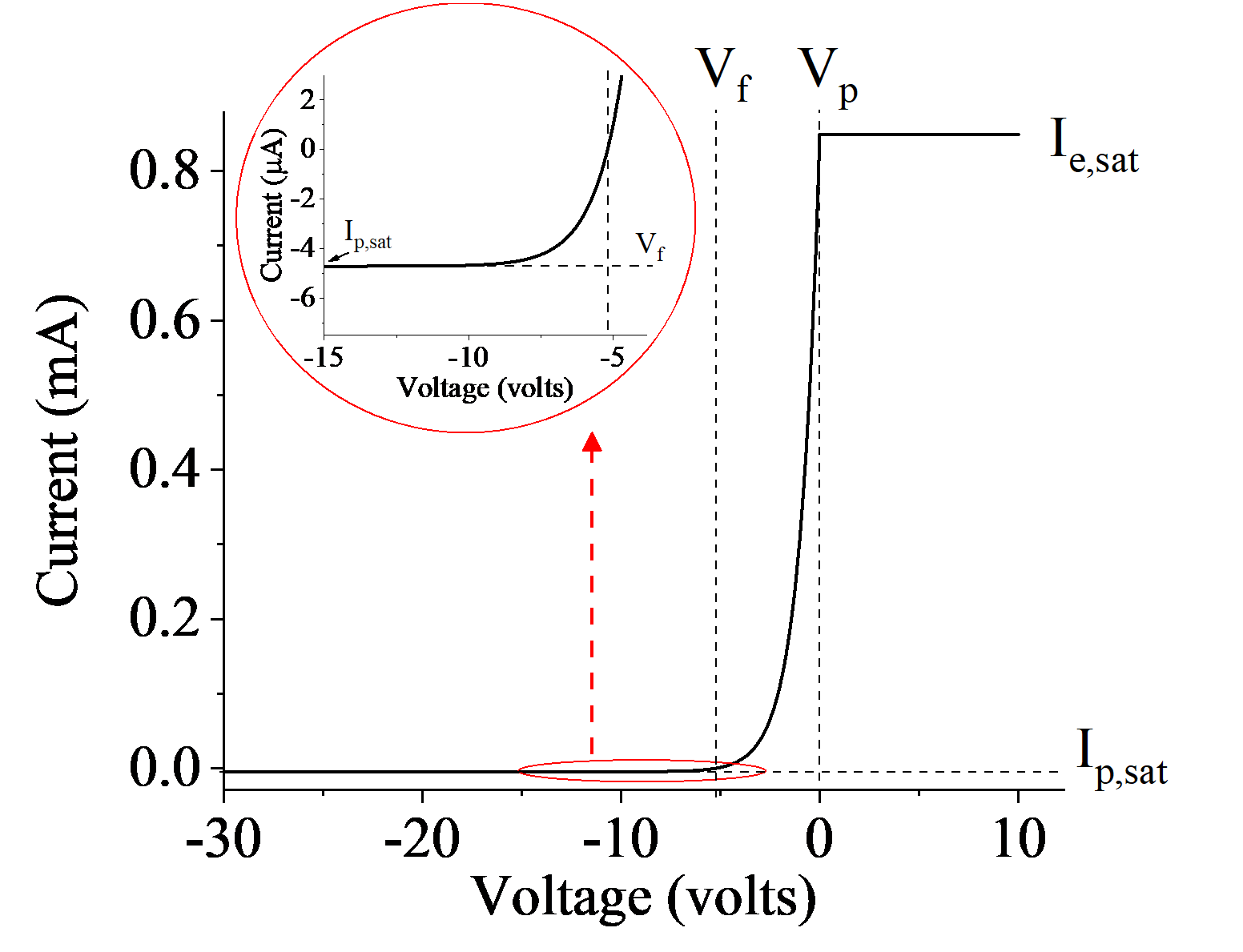}
    \caption{\label{IVthr} Plot of an ideal IV characteristic for $T_e$ = 1 eV and $n_{p0}$ $\approx$ $n_{e0}$ =$10^{16} m^{-3}$. $V_f$ = floating potential,  $V_p$ = plasma potential, $I_{p,sat}$ = positive ion saturation current, $I_{e,sat}$ = electron saturation current.}
  
\end{figure*}

A Langmuir probe (LP) is simply a bare tungsten or molybdenum wire which is introduced inside a discharge tube to collect charge particles from the plasma. As the probe is electrically biased relative to a reference electrode, it draws a current which varies exponentially between the electron saturation region at the plasma potential and the positive ion saturation current corresponding to a sufficiently large negative bias on the probe (figure~\ref{IVthr}). In electronegative plasmas, both electrons and negative ions contributes to the saturation probe current at $V_p$. Ideally for a planar, non-emitting probe, the electron to positive ion saturation current ratio can be expressed by: 

\begin{eqnarray}
R =\frac{I_{-,sat}}{I_{+,sat}}=\frac{e(0.25n_{v0}v_v+0.25n_{e0}v_e)A_{probe}}{en_{ps}u_BA_{sheath}} \label{thr1}
\end{eqnarray}

In the above equation, $n_{v0}$ and $n_{e0}$ denotes the negative ion density and the electron density in the bulk. Their thermal speeds are denoted by $v_v$ and $v_e$ respectively. The positive ion saturation current, $I_{+,sat}=en_{ps}u_BA_{sheath}$  , wherein $ n_{ps}$ is the positive ion density at the sheath edge has been assumed rather than the bulk plasma density ($n_{po}$). This is because the positive ion density falls monotonically from the bulk plasma to the sheath edge due to the potential drop of the order of $\eta_s \sim \frac{kT_e}{e}$ existing inside the pre-sheath. This is however not the case when the probe is biased at the plasma potential $V_p$ to collect the electrons / negative ions. 
 The plasma density at the sheath edge follows a Boltzmann distribution, $n_s \sim n_0exp(\frac{e\eta_s}{kT_e})$. Therefore for the case of a pure electro-positive plasma, the sheath edge density is $n_s \sim 0.61n_{p0}$. However in the case of electro-negative plasma, the potential drop inside the pre-sheath can vary according to the electronegativity parameters, $\alpha=\frac{n_{v0}}{n_{e0}}$ or $\alpha_s=\frac{n_{vs}}{n_{es}}$ and electron to negative ion temperature ratio, $\gamma_v=\frac{T_e}{T_v}$ . This has a direct influence on the sheath edge density and accordingly, the electronegative parameters around the probe will be modified. Assuming the charge species follows Boltzmann relation in the presheath having potential fall $\eta_s$, the electronegativity parameter at the sheath edge, $\alpha_s$ can be related to the electronegativity parameter in the bulk, $\alpha$ as follows

\begin{eqnarray}
\alpha_s=\alpha exp(-\eta_s(\gamma_v-1)) \label{thr2}
\end{eqnarray}

      It is also important to note that for the case of cylindrical probes, the sheath width around the probe surface increases with application of probe bias; hence the current collection surface including the sheath $A_{sheath}$  increases for the case of positive ions, in which the flux of positive ions remain conserved across the sheath. On the other hand, during application of positive bias, the sheath width is vanishingly small, hence the electrons and negative ions flux enters the sheath with their thermal velocity according to a Boltzmann distribution and gets collected at the probe surface have area $A_{probe}$. If $\phi_B$ is the relative bias between the probe and the plasma, the a parameter k($\phi_B$) can be defined as the relative ratio of the sheath to the probe surface area, 

\begin{eqnarray}
k(\phi_B) = \frac{A_s}{A_p} \label{thr3}
\end{eqnarray}
and the positive ion speed at the sheath edge ($u_B$), also known as bohm velocity, as

\begin{eqnarray}
u_B=\sqrt{\frac{eT_e}{M}} \sqrt{\frac{1+\alpha_s}{1+\gamma_v\alpha_s}}=\sqrt{\frac{eT_e}{M}} \sqrt{\frac{1+\alpha exp(-\eta_s(\gamma_v-1))}{1+\gamma_v\alpha exp(-\eta_s(\gamma_v-1))}} \label{thr4}
\end{eqnarray}

In equation-\ref{thr4}, M is the positive ion mass and  $u_B$ as the Bohm velocity \cite{Sheridan1999,Pandey2017}.Assuming the positive ions to be at rest at the pre-sheath boundary, the simple energy balance equation can be applied to determine the pre-sheath potential drop ($\eta_s$) in terms of $\alpha_s$ and $\gamma_v=\frac{T_e}{T_v}$   as 

\begin{eqnarray}
\eta_s=\frac{T_e}{2}\frac{1+\alpha_s}{1+\gamma_v\alpha_s}=\frac{T_e}{2}\frac{1+\alpha exp(-\eta_s(\gamma_v-1))}{1+\gamma_v\alpha exp(-\eta_s(\gamma_v-1))} \label{thr5}
\end{eqnarray}

Substituting $\eta_s$ from equation-\ref{thr5} in equation-\ref{thr3}, the negative ion parameter at the sheath-presheath edge ($\alpha_s$) w.r.t. plasma-presheath edge ($\alpha$), can be expressed by

\begin{eqnarray}
\alpha=\alpha_sexp(\frac{(\gamma_v-1)(1+\alpha_s)}{1+\gamma_v\alpha_s}) \label{thr6}
\end{eqnarray}

Using equation-\ref{thr2} to \ref{thr6} in equation-\ref{thr1}, the ratio of saturation currents to the probe can now be expressed as; 

\begin{eqnarray}
R=\frac{0.25e\alpha v_{thv}+0.25e v_{the}}{k(\phi_B)e^{\eta_s}(1+\alpha)e \sqrt{\frac{eT_e}{M}} \sqrt{\frac{e^{\eta_s}+\alpha e^{\eta_{\gamma_v \eta_s}}}{e^{\eta_s}+\alpha \gamma_v e^{\eta_{\gamma_v \eta_s}}}}} \label{thr7}
\end{eqnarray}

where $\frac{n_{p0}}{n_{e0}=1+\alpha}$  and can be re-arranged into, 

\begin{eqnarray}
\fl (R^2 - \frac{\mu_{pv}}{2\pi e^{-2\eta_s}k^2}) \alpha^3 exp(-(\gamma_v-1)\eta_s)  + [ 2R^2 exp(-(\gamma_v-1)\eta_s)+R^2\nonumber\\-\frac{\mu_{pv}}{2 \pi e^{-2\eta_s}k^2\gamma_v} -2\frac{\sqrt{\mu_{pv}\mu_{pr}\gamma_v}}{2 \pi e^{-2\eta_s}k^2}exp(-(\gamma_v-1)\eta_s)] \alpha^2 \nonumber\\+ [2R^2exp(-(\gamma_v-1)\eta_s)+R^2- \frac{2}{2 \pi e^{-2\eta_s}k^2} \sqrt{\frac{\mu_{pv}\mu_pe}{\gamma_v}}\nonumber\\ - \frac{\gamma_v\mu_{pe}}{2 \pi e^{-2\eta_s}k^2} exp(-(\gamma_v-1)\eta_s)] \alpha + R^2 - \frac{\mu_{pe}}{2 \pi e^{-2\eta_s}k^2}=0 \label{thr8}
\end{eqnarray}

where $\mu_{pv}=\frac{M}{m_v}$ and  $\mu_{pe}=\frac{M}{m_e}$. 

The above equation basically relates the saturation current ratio R to the bulk electronegativity parameter, $\alpha$. The remaining parameters  $\gamma_v$ , $\eta_s$ and the sheath area correction factor k around the probe are usually remains constant; but they can significantly affect the negative ion estimation. For instance the parameter $\eta_s$ which represents the potential drop in the pre-sheath is a function of negative ion density. In some cases, $\eta_s$ has been assumed to be 0.5 like in the case of electro-positive plasma \cite{Sheridan1999,Bredin2014,Bowes2014,Pandey2017}; this approximation may only be valid for weakly electronegative discharges. The other possible errors arise during the estimation of positive ion saturation current due to uncertainty in defining the actual sheath radius. The sheath radius is again a function of bulk electronegativity parameter, which needs to be taken in to consideration during the calculation of positive ion saturation current. In many cases, the probe collection area had been approximated to the probe radius which is generally valid under thin sheath approximation \cite{Sheridan1999,Bredin2014}. However the sheath radius around the cylindrical wire probe is also a function of negative ion density. Finally negative ion temperature ratio $\gamma_v$ is required to be substituted in equation-\ref{thr8} to express $\alpha$ in terms of R; where the $\gamma_v$ is usually assigned as a constant parameter, in which negative ion temperature, $T_v$ is typically assumed higher than the temperatures of the gas neutrals ($T_v$ = 0.05 eV).

\section{Corrections due to pre-sheath and sheath around cylindrical wire probe}\label{thrcorr}

In equation-\ref{thr8}, several parameters like the saturation current ratio factor R, the sheath-edge potential $\eta_s$, temperature ratio $\gamma_v$ and the sheath area k are required for finding the $\alpha$. Uncertainties in these quantities can lead to possible errors in the estimation of negative ion parameters. The individual cases are briefly discussed as follows: 

\subsection{Sheath edge potential/ Presheath fall ($\eta_s$) }

\begin{figure*}[bbp]
  \centering
     \includegraphics[width=0.5\textwidth]{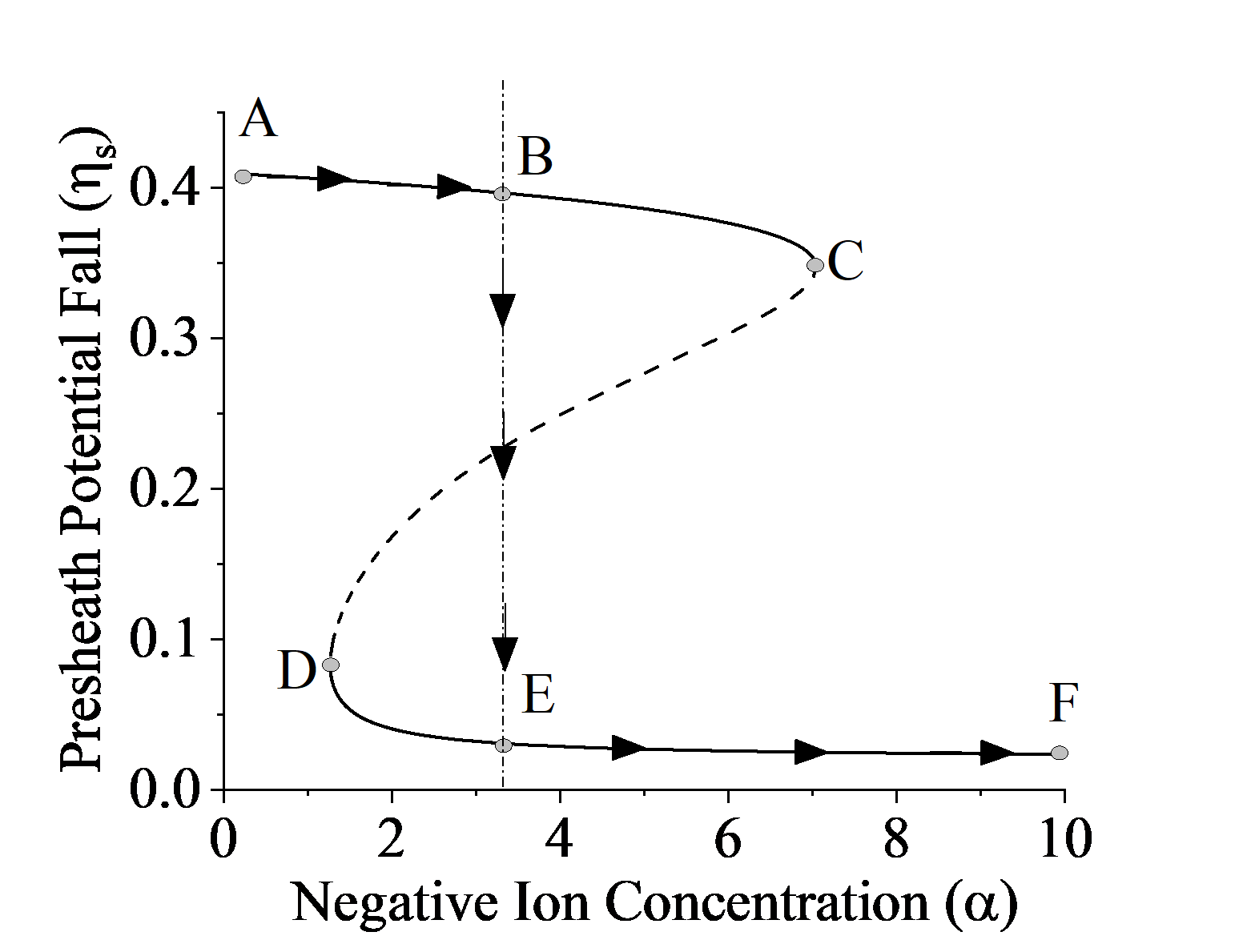}
    \caption{\label{etavsalpha} Plot of presheath potential fall versus negative ion density at presheath edge for $\gamma_v$ = 40. Vertical line is at $\alpha$ = 3.23. For $\alpha$ = D to C, the solution shows a triple value solution. The arrows A-B-E-F shows the physical feasible solution.}
  
\end{figure*}

The pre-sheath potential fall can be obtained from equation-\ref{thr5}, which is a function of $\alpha$ and $\gamma_v$. As plotted in figure-\ref{etavsalpha}, the analytical solution for a range of $\alpha$ shows that the potential fall in the pre-sheath is double valued if $\gamma_v \geq 5+\sqrt{24}$. This condition is usually satisfied in low pressure laboratory plasma devices. Since the fall in pre-sheath potential directly influences the positive ion density at the sheath edge, its omission could give rise to errors in the negative ion density determination. Therefore it is important to determine the $\eta_s$, which has a direct influence on the positive ion density at the sheath edge.

\begin{figure*}[!hbp]
  \centering
     \includegraphics[width=0.5\textwidth]{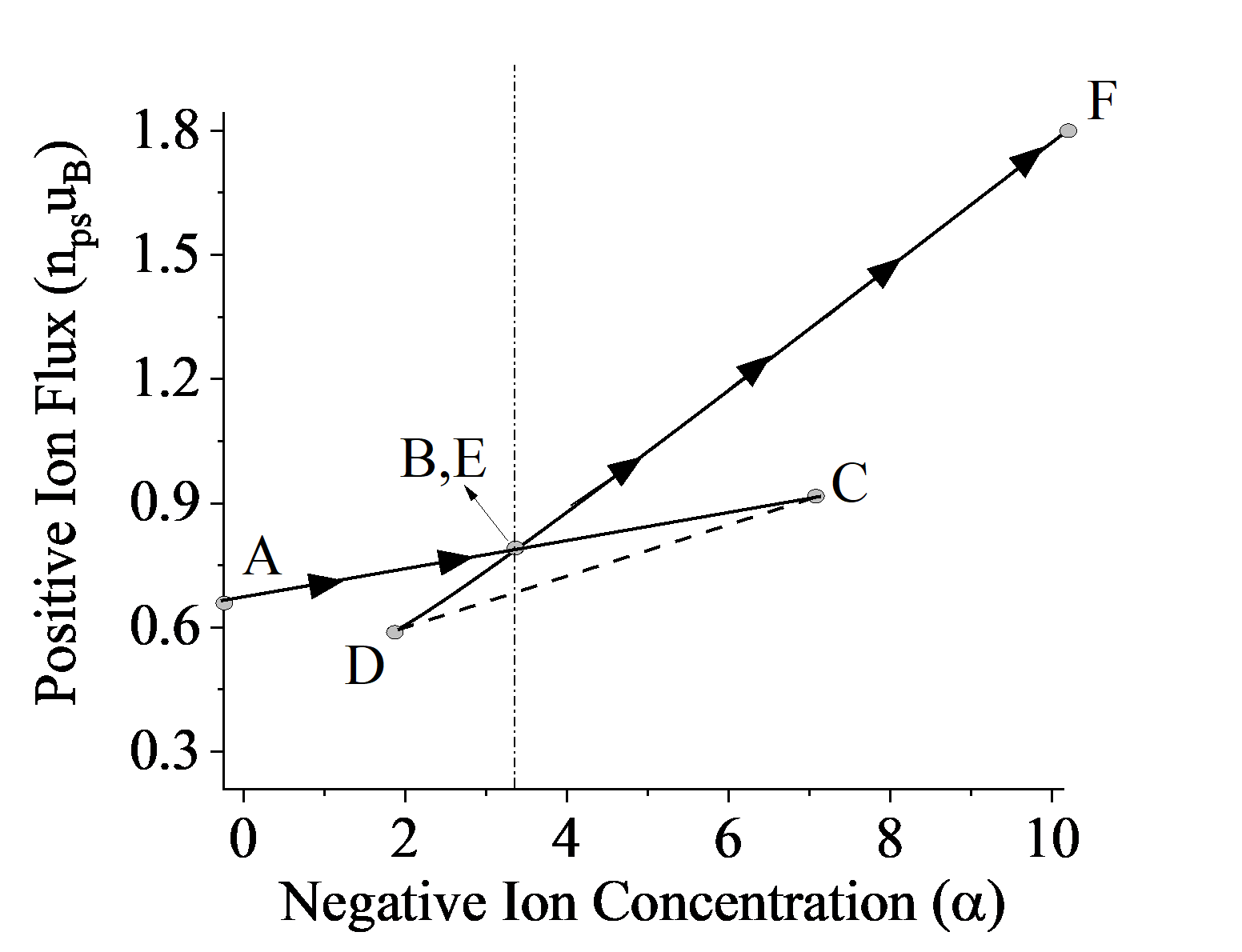}
    \caption{\label{fluxvsalpha} Plot of positive ion flux at sheath edge versus negative ion concentration in plasma for $\gamma_v$ = 40. Vertical line is at $\alpha$ = 3.23. For $\alpha$ = D to C, the solution shows a triple value solution. The arrows A-B-F shows the maximum flux. }
  
\end{figure*}

In conjunction to above the positive ion flux collected by the probe is plotted in figure-\ref{fluxvsalpha}. The straight lines corresponds to the possible values of positive ion flux with regard to the sheath edge density, which is dependent over the sheath edge potential, whereas the positive ion speed is constant. Sheridan et al. \cite{Sheridan1999} proposed that the correct criterion for calculating the flux at the sheath edge should consider the larger predicted value of positive ion flux. The above findings were also reported by Amemiya et al. \cite{Amemiya1999} but following a totally different approach by taking the positive ion density in terms of collected current rather than using the energy equation for solving Poisson’s equation. 

In figure-\ref{etavsalpha} and figure-\ref{fluxvsalpha}, the vertical line demarks the two possible regimes in which the pre-sheath potential drop could change abruptly. For a particular value of  $\gamma_v$, the value of the presheath potential can be evaluated corresponding to a particular value of $\alpha$ according to the arrows shown in figure-\ref{etavsalpha}.

\subsection{Positive ion sheath}

The positive ion sheath is formed by providing sufficiently high negative potential on the probe to repel the tail electrons arriving at the probe surface; therefore the probe current is merely constituted of ions. Theoretically, the sheath width can be calculated by solving the Poisson’s equation with appropriate boundary conditions \cite{Lieberman2005}. The Poisson’s equation for a cylindrical probe sheath can be expressed as;

\begin{eqnarray}
r \frac{\partial^2 \phi(r)}{\partial r^2} + \frac{\partial \phi(r)}{\partial r} = \frac{er}{\epsilon_0}(n_e(r)-n_v(r)-n_p(r)) \label{thrcorr1}
\end{eqnarray}

where $\epsilon_0$ is the dielectric constant, $\phi$ is the potential,$n_e$, $n_v$ and $n_p$ are the electrons, negative ions and positive ions densities respectively. Using the continuity equation,

\begin{eqnarray}
\frac{\partial rn_pv_p}{\partial r} = 0 \label{thrcorr2}
\end{eqnarray}

the momentum equation;

\begin{eqnarray}
\frac{\partial rn_pv_p^2}{\partial r} = -\frac{e}{M}n_pr \frac{\partial \phi(r)}{\partial r} \label{thrcorr3}
\end{eqnarray}

and the boundary conditions at the sheath edge, s is given by;

\begin{eqnarray}
\phi(s) \approx 0, n_p(s)=n_{ps}, v_p(s)=u_B, \phi(0) = \phi_{bias}, \phi'(s) \approx 0, \label{thrcorr4}
\end{eqnarray}

the positive ion density inside the sheath is expressed as;

\begin{eqnarray}
n_p(r)=\frac{sn_{ps}}{r}(1-\frac{2e\phi(r)}{M u_B^2})^{-0.5} \label{thrcorr5}
\end{eqnarray}

The electrons and negative ions are considered to have a Boltzmann distribution, which is generally true in the case of low-pressure plasma discharges. Hence the corresponding densities inside the sheath can be expressed as;

\begin{eqnarray}
n_{v,e}(r)=n_{v0,e0}exp(\frac{e\phi(r)}{k_BT_{v,e}}) \label{thrcorr6}
\end{eqnarray}

\begin{figure*}[bbp]
  \centering
     \includegraphics[width=0.65\textwidth]{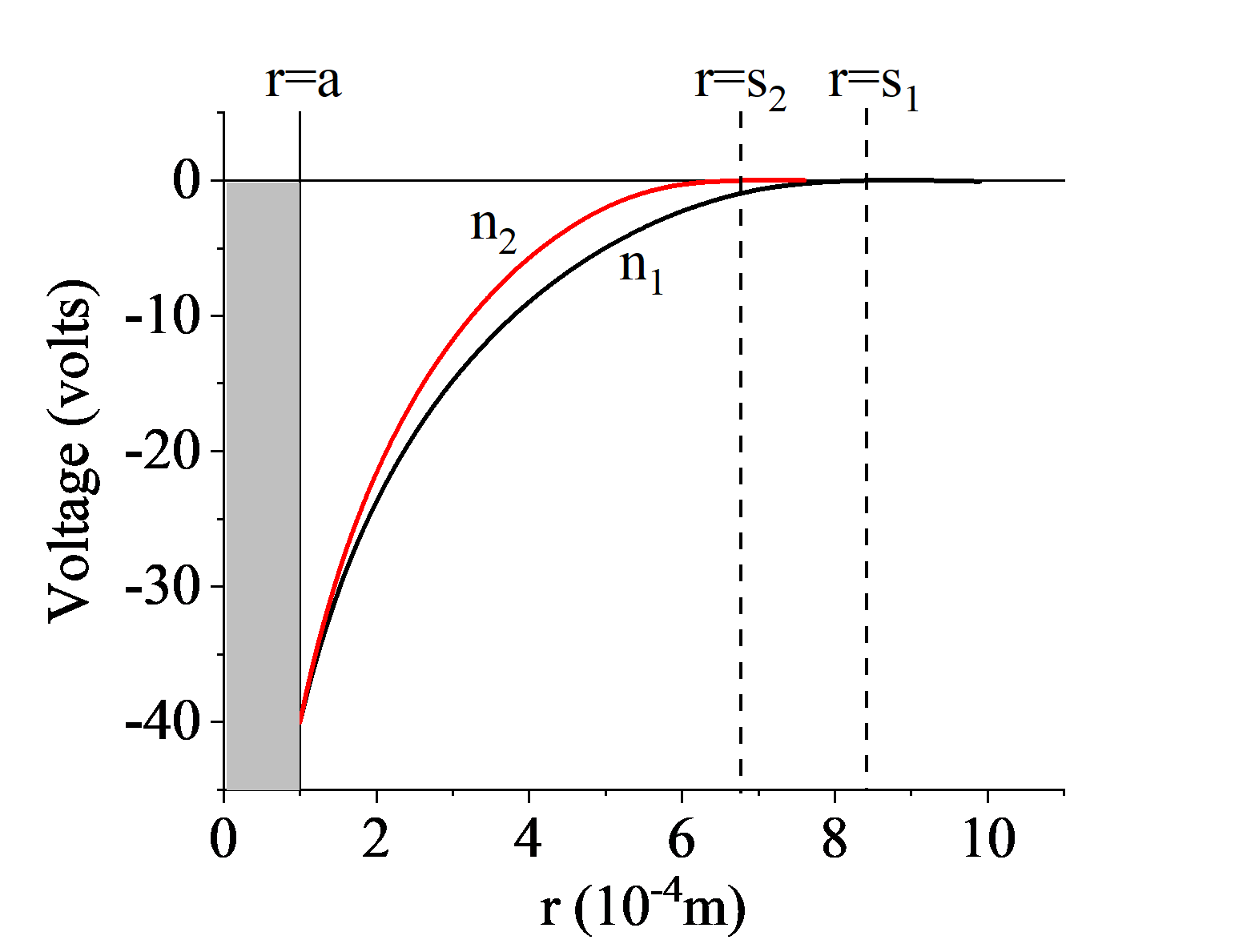}
    \caption{\label{sheathpot} Plot of voltage drop inside the sheath, a = probe radius, $n_1$ = 1 $\times$ $10^{16} m^{-3}$, $n_2$ = 2 $\times$ $10^{16} m^{-3}$, $s_1$ and $s_2$ are the corresponding sheath width for fix probe voltage of -40 Volts and electron temperature of 1 eV.}
  
\end{figure*}

Using equations-\ref{thrcorr4}, \ref{thrcorr5} and \ref{thrcorr6} in equation-\ref{thrcorr1}, the potential inside the probe sheath can be found, hence the spatial extent of the sheath potential as function of probe bias, $\phi(r)$ can provide the information of the sheath width.

The sheath width is required to calculate the saturation positive ion current collected by the probe. However the sheath width is dependent on the sheath edge density, which requires the prior knowledge of $\alpha$ and positive ion density at the sheath edge $n_{ps}$. In figure-\ref{sheathpot}, the dependence of sheath edge density on the spatial potential distribution inside the sheath is shown. It can be seen that the sheath width reduces with an increase in the sheath edge density.

In one of our papers, the sheath width around a cylindrical probe immersed in electronegative oxygen plasma had been determined by applying a biased hairpin probe technique. This method can be applied to compensate for the sheath width around the cylindrical wire to estimate the positive ion density at the sheath edge.

\section{Errors in the estimation of electron and positive ion saturation current}\label{experr}

\subsection{The electron saturation region}

\begin{figure*}[bbp]
  \centering
     \includegraphics[width=0.5\textwidth]{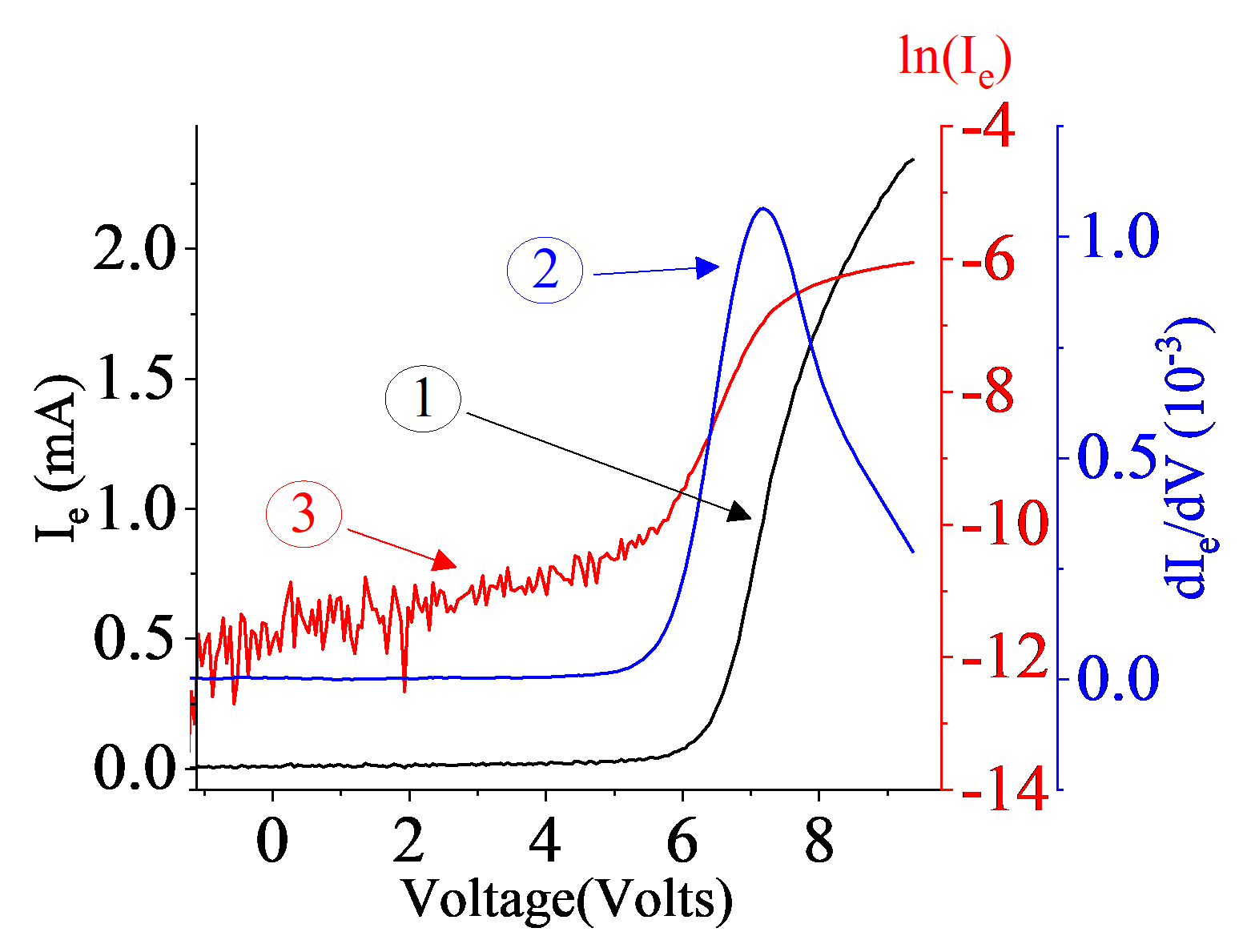}
    \caption{\label{IVexp} Plot of a typical IV characteristic including 1st derivative and natural log of current.(1) Electron current versus voltage, (2) first derivative of electron current versus voltage, (3) natural log of electron saturation current versus voltage.}
  
\end{figure*}

Figure-\ref{IVexp} presents a typical Langmuir probe characteristics obtained in a DC discharge produced using a constricted anode and a pair of annular planar cathodes. From this characteristic, the plasma potential is determined using the first derivative of the probe current and then electron saturation current corresponding to this plasma potential. The electron temperature is found from the semi-log plot of I-V characteristics. 

In a Maxwellian argon plasma, the cylindrical probe assumes a floating potentia, $V_f$ relative to the plasma potential $V_p$, according to the following relation \cite{lieberman2005principles}:

\begin{eqnarray}
V_p-V_f=5.2T_e \label{experr1}
\end{eqnarray}

\begin{figure*}[bbp]
  \centering
     \includegraphics[width=0.5\textwidth]{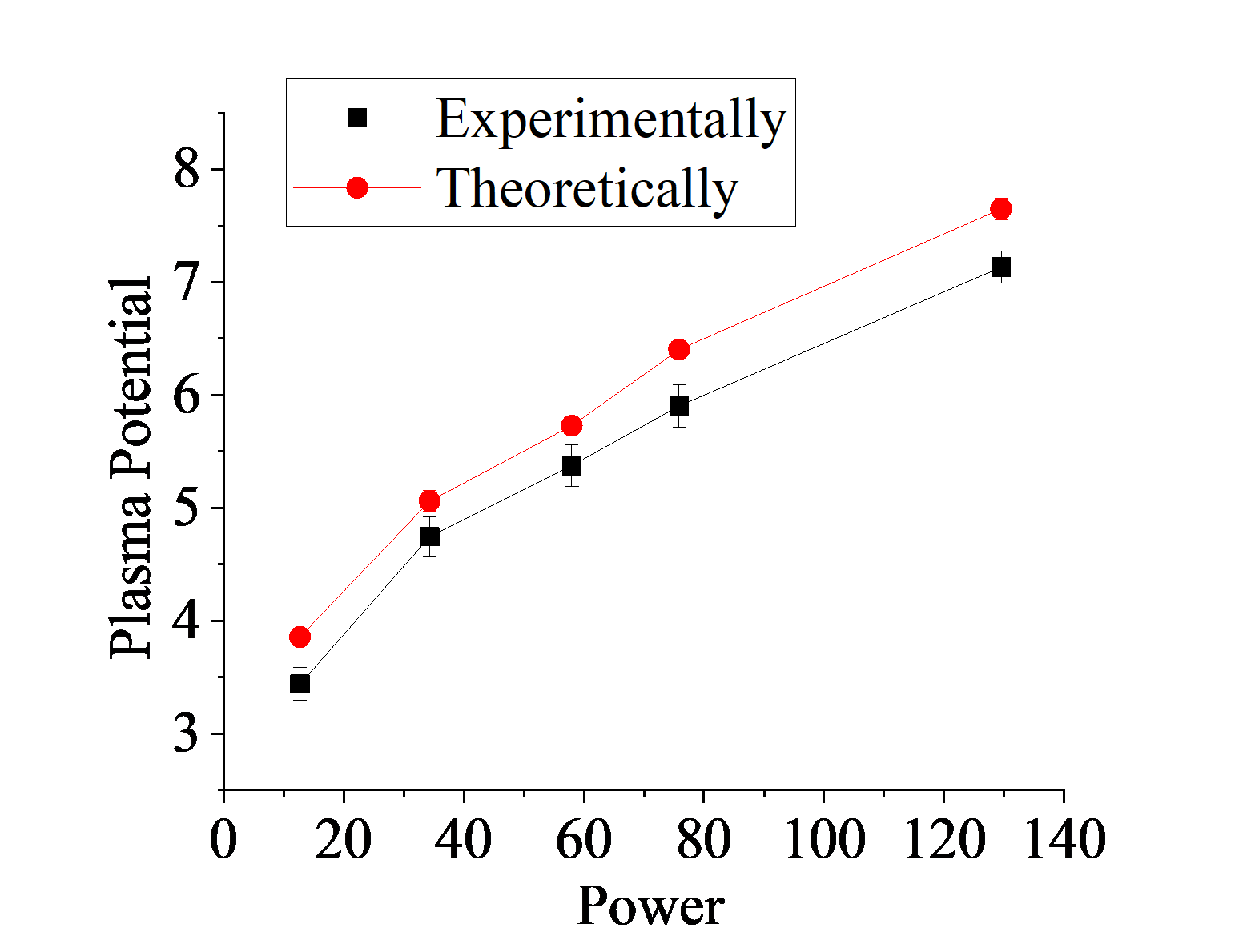}
    \caption{\label{plasmapot} Plot of Plasma Potential calculated from the peak of the 1st derivative of $I_e$ w.r.t. V and from $V_p$ = $V_f$ + 5.2 $T_e$ versus Discharge Power.}
  
\end{figure*}

\begin{figure*}[!hbp]
  \centering
     \includegraphics[width=0.5\textwidth]{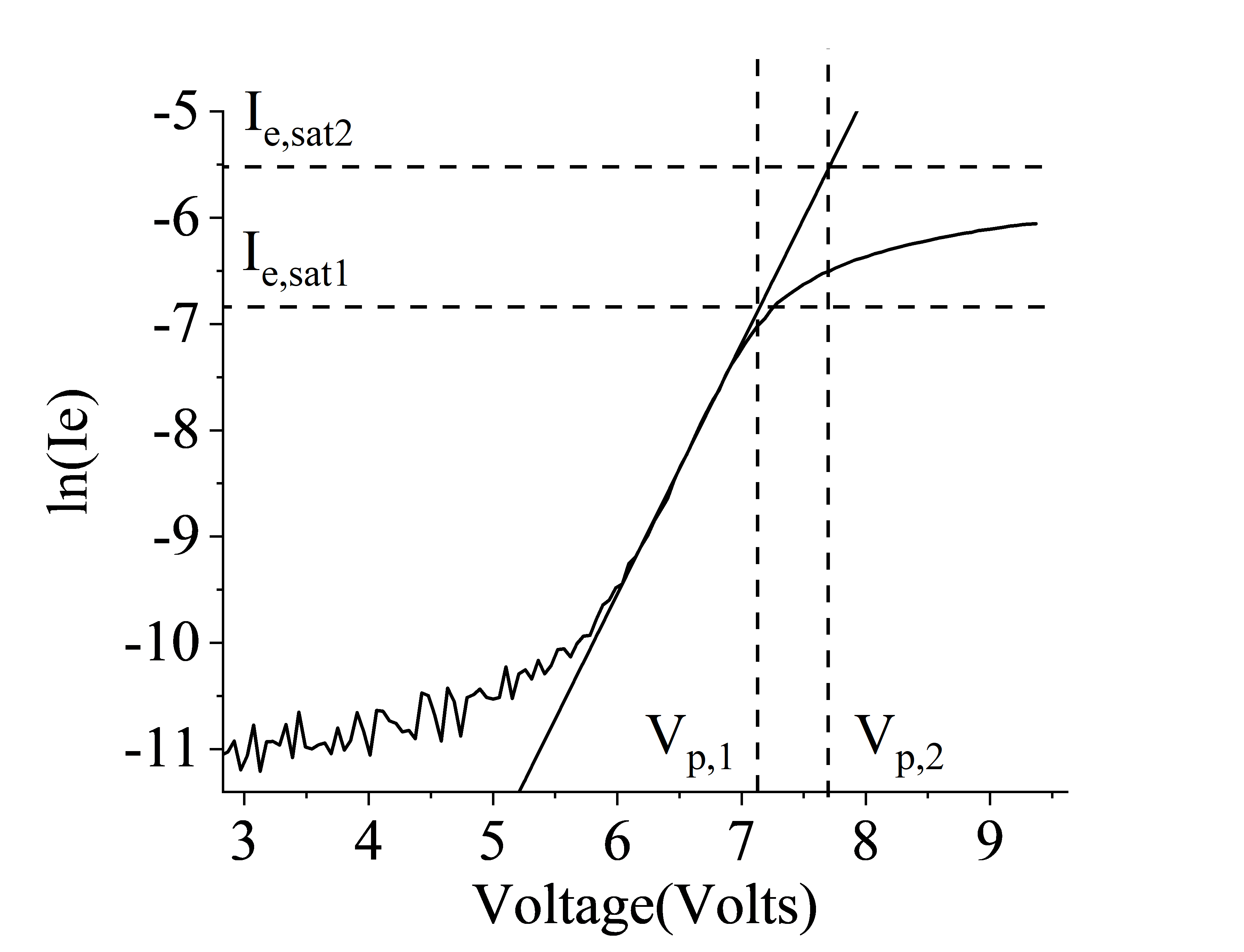}
    \caption{\label{lnIe} The straight line fit is extrapolated upto the plasma potential. In the vertical dash lines, $V_{p,1}$ and $V_{p,2}$ are the plasma potential from the 1st derivative of I-V and from $V_p$ = $V_f$ + 5.2 $T_e$ respectively. The  $I_{e,sat1}$ and $I_{e,sat2}$ in the horizontal dash lines are the corresponding electron saturation current.}
  
\end{figure*}

\begin{figure*}[tbp]
  \centering
     \includegraphics[width=0.5\textwidth]{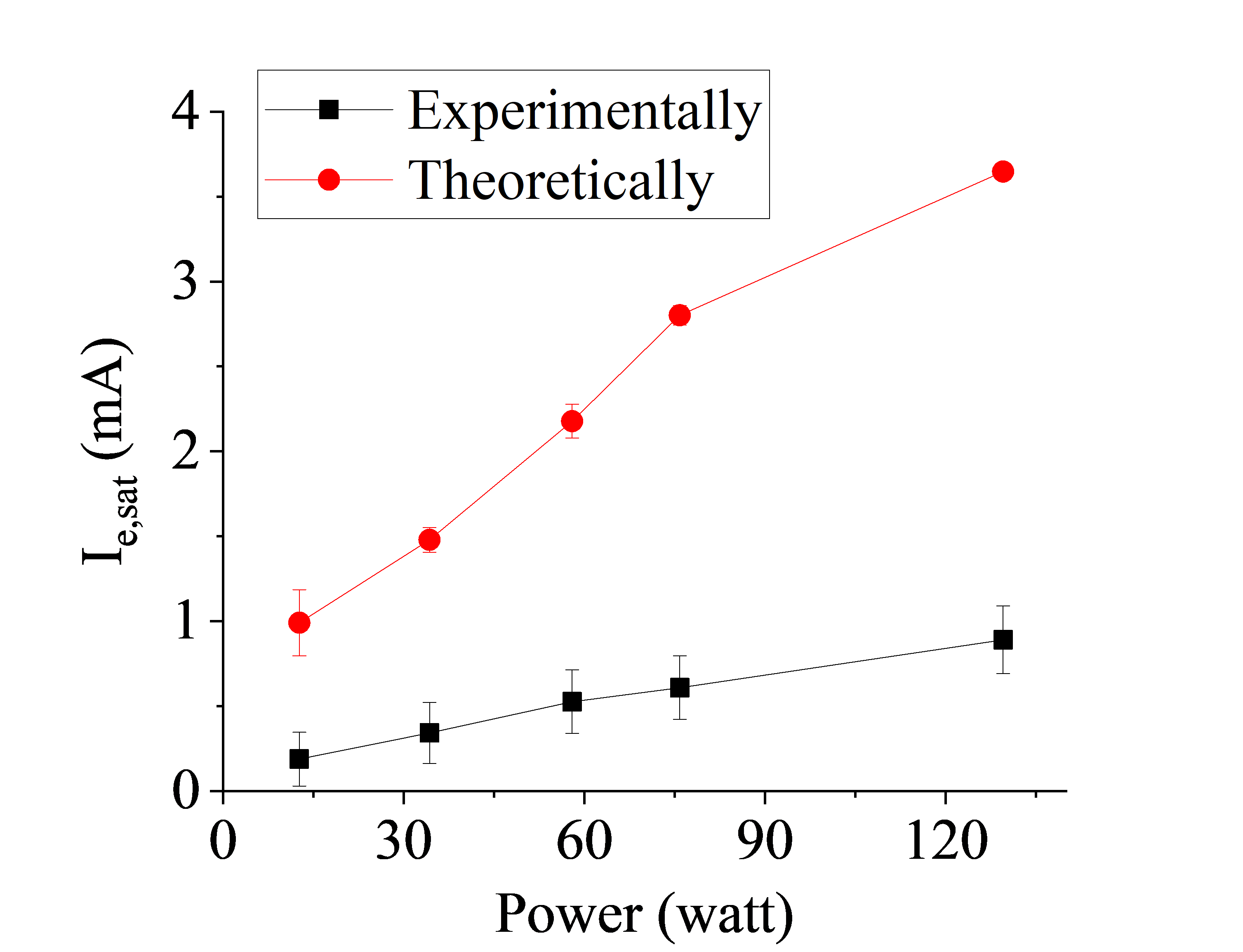}
    \caption{\label{Iesat} Electron saturation current calculated from the peak of the 1st derivative of $I_e$ w.r.t. V and from $V_p$ = $V_f$ + 5.2 $T_e$ versus Discharge Power.}
  
\end{figure*}

From this characteristics (figure-\ref{IVexp}) the electron temperature is found to be $T_e$ = 1.4 eV, floating potential $V_f$ = -4.08 V, therefore according to equation-\ref{experr1} it provides a theoretical value of $V_p$ = 3.9 V , which is higher than the $V_p$ obtained by applying the first derivative method in figure-\ref{IVexp} by a factor of 4-7 as compared to floating potential method. This factor is seems to be consitent with applied power as shown in figure-\ref{plasmapot}. Corresponds to this theoretical values of $V_p$, the electron saturation current can be estimated by extrapolating the straight line in the natural log plot of electron current as shown in figure-\ref{lnIe}. Clearly a small change in this value can drastically affect the estimation of electron saturation current beacause of the associated nature of exponential. This is shown in figure-\ref{Iesat}. 

There are various factors that can lead to the underestimation in plasma potential attained by a cylindrical probe. The rounding of the knee region is sometime difficult to locate the actual position of $V_p$ relative to the applied probe bias. As the probe collects electrons, an equivalent amount of positive ion current must reach the reference grounded electrode. If the reference electrode is far away from the plasma or it is improperly grounded, it could lead to a virtual resistance in the current path \cite{Bhuva2019}, which limits the electron saturation current to attain its maximum value. Furthermore, external magnetic fields can also reduce the electron saturation current. With this underestimated electron saturation current, the negative ion parameters estimation is directly influences using the saturation current ratio method. Hence the floating potential method based on emissive probe can be a reliable means to estimate the plasma potential.

\subsection{The positive ion saturation region}

Theoretically the positive ion saturation current is expressed as the product $I_{i,sat}=en_{ps} A_s u_B$ wherein the sheath edge density and the Bohm speed is constant; for a planar sheath the area can be assumed as the probe area, however for a cylindrical wire the current collection area increases with the probe bias, as a result the net current due to positive ions increases.

\begin{figure*}[tbp]
  \centering
     \includegraphics[width=0.5\textwidth]{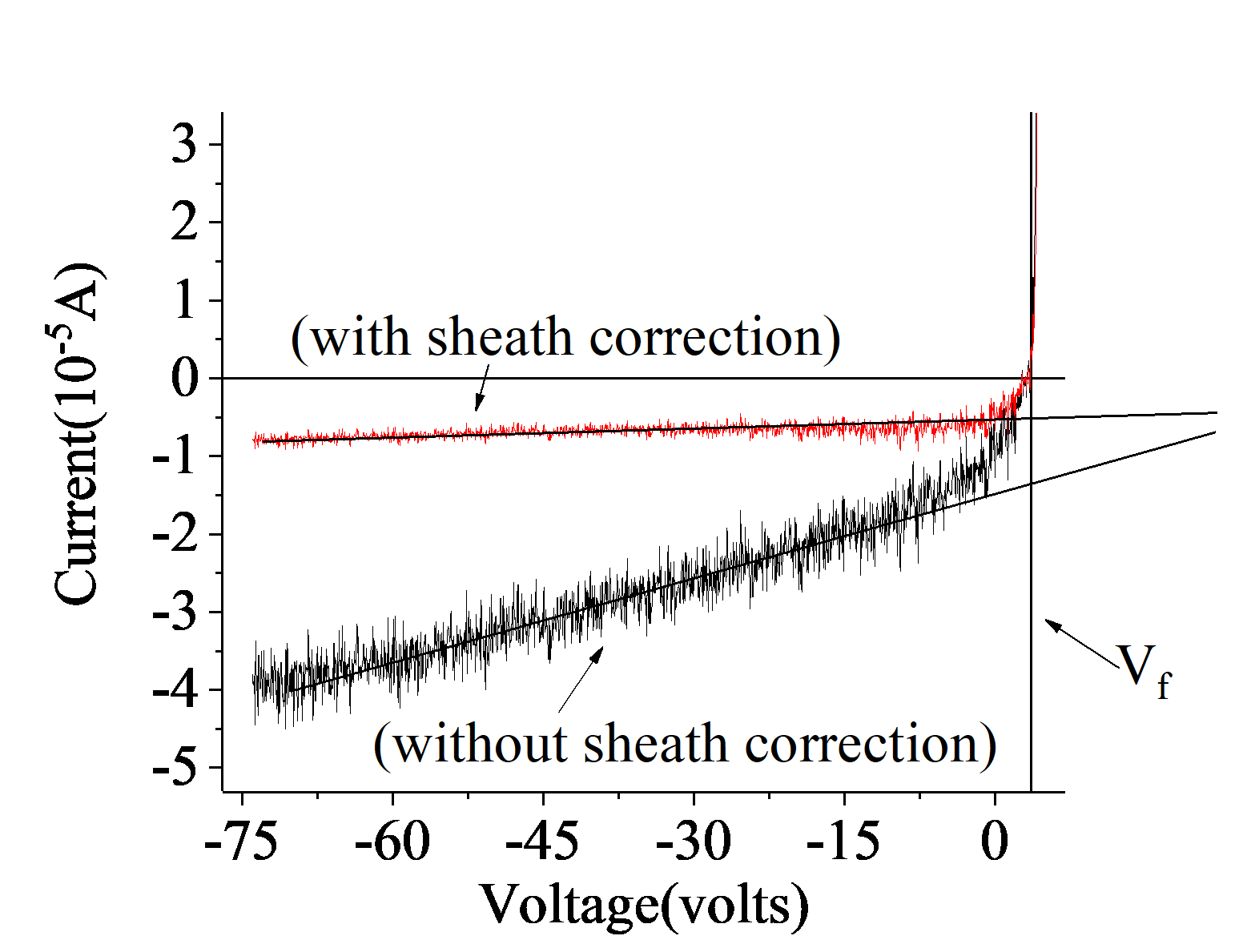}
    \caption{\label{IVcorr} Plot of current collection to probe without any correction and with sheath correction versus Voltage.}
  
\end{figure*}

\begin{figure*}[tbp]
  \centering
     \includegraphics[width=0.5\textwidth]{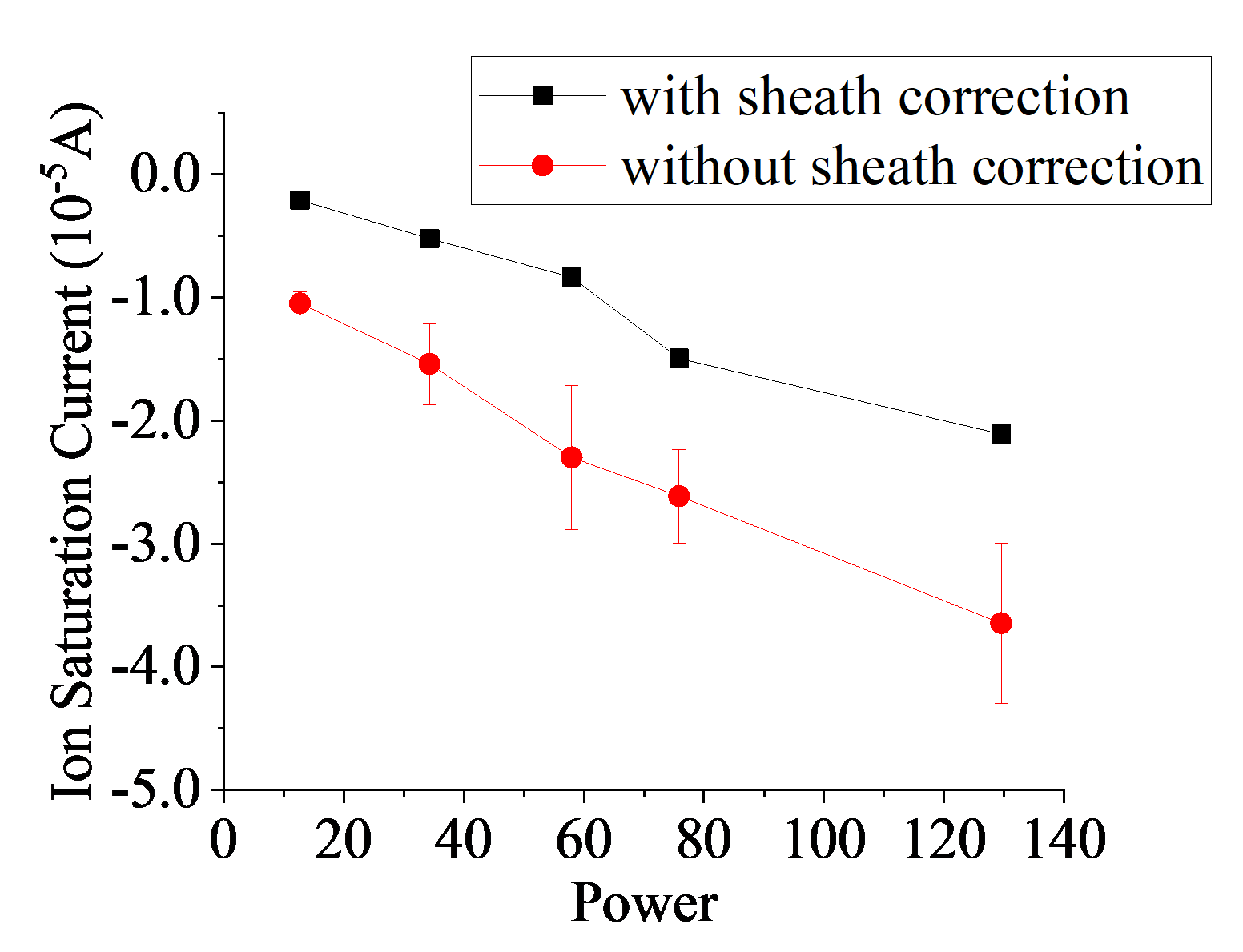}
    \caption{\label{Iisat} Plot of ion saturation current without any correction and with sheath correction versus Voltage.}
  
\end{figure*}

In figure-\ref{IVcorr}, the ion saturation current is found to increase almost linearly with the application of negative probe bias. This increase in current is caused due to an expansion in the sheath around the cylindrical wire. Pandey et al. \cite{Pandey2020} applied a resonance hairpin probe to estimate the sheath width expansion around a cylindrical wire probe of same diameter. With the knowledge of the sheath width due to negative bias on the probe, it is possible to remove the expansion of the ion saturation current as shown by plotting the revised ion saturation current. To estimate the ion saturation current, Chen et al. \cite{Chen2002} suggested that it is convenient to consider the ion current corresponding at the floating potential. It can be clearly seen in figure-\ref{IVcorr} that the value of ion saturation current at the floating potential, with/ without including the sheath correction differs significantly. The corrected value of ion saturation current for a range of power is also shown in figure-\ref{Iisat}.


\section{Experimental results and Discussion}\label{ExpRes}

\subsection{The experimental setup}

In order to apply the saturation current ratio method to obtain negative ion density, a DC discharge was created using a constricted anode and parallel plate cathodes, similar to the one used by Pandey at al. \cite{Pandey2017}. Both electropositive argon and electronegative oxygen plasma was created. In this setup, the plasma density can be varied in the range of $10^{16}$ -$10^{17}$ $m^{-3}$ for a range of pressure between 1-3 Pa and by varying the discharge power from 10 to 150 Watt.

\subsection{Plasma parameters measurement using LP}

For the measurement of plasma parameters a cylindrical Langmuir probe (LP) having wire radius 0.25 mm and length 5.0 mm was introduced in the centre of the discharge. An I(V) characteristics was obtained by biasing the probe from the ion saturation region corresponding to bias voltage in excess of – 75 V to the electron saturation region at the plasma potential, which was typically at +10 V relative to the grounded chamber. A typical characteristics of the I(V) plot is already shown in figure-\ref{IVexp} from which the electron temperature ($T_e$) can be calculated by plotting a semi-log plot of the electron saturation current according to the standard probe theory discussed elsewhere \cite{Merlino2007}.

\begin{figure*}[!hbp]
  \centering
     \includegraphics[width=0.5\textwidth]{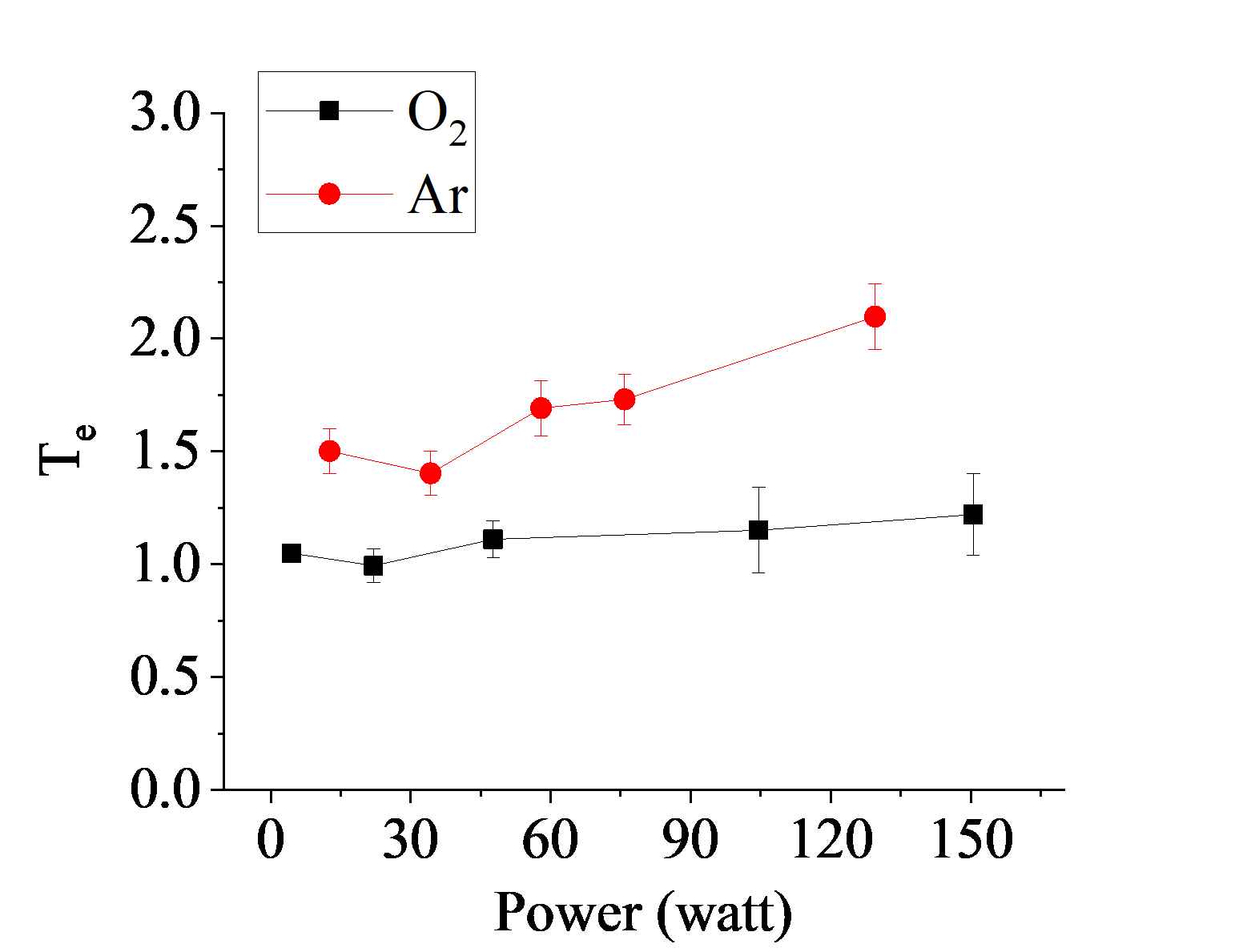}
    \caption{\label{Temp} Plot of electron temperature versus Discharge power for argon and oxygen plasma.}
  
\end{figure*}

       In figure-\ref{Temp}, typical values of $T_e$ obtained in argon and in oxygen plasma are plotted as a function of discharge power. The variation in $T_e$ is found to be almost constant for the case of oxygen, whereas it increases slightly for argon, however the variation remain within 25\% over the entire range.  

The electron temperature has a small contribution to the overall electron saturation current, but as discussed in section-\ref{experr}, the error in plasma potential due to rounding of the electron saturation current is quite significant. Hence the plasma potential is determined from the floating potential corresponding to the electron temperatures obtained in figure-\ref{Temp}. It is important to mention that for the case of oxygen plasma, the mass ratio between the electron and positive oxygen ion has been considered for estimating the plasma potential which is $V_p=V_f+5.07 T_e$; whereas for the case of argon plasma this difference between the $V_p$ and$ V_f$ is $5.2 T_e$.

\begin{figure*}[!hbp]
  \centering
     \includegraphics[width=0.5\textwidth]{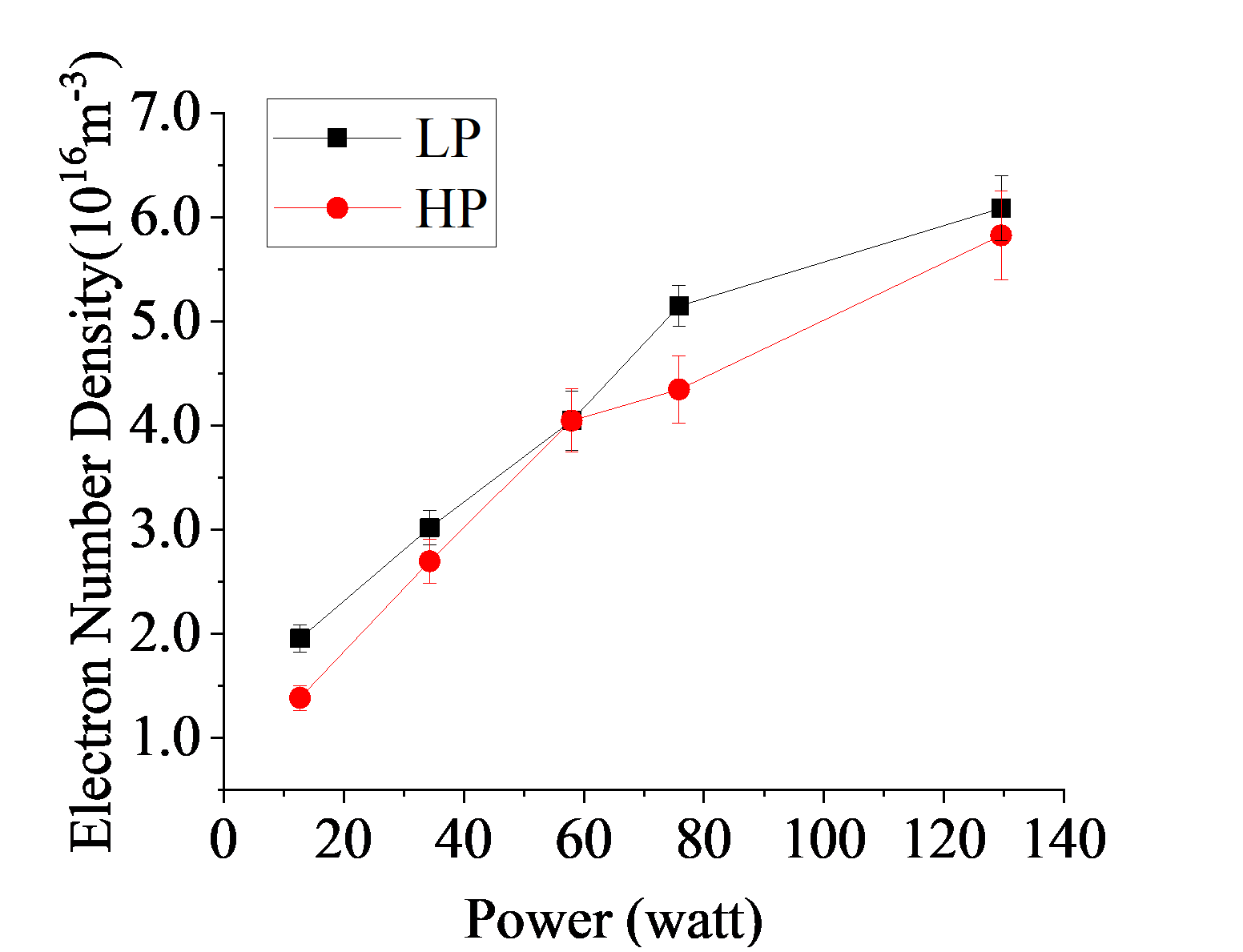}
    \caption{\label{LPHP} Plot of electron number density calculated using HP and from electron saturation of LP versus discharge power.}
  
\end{figure*}

The plots of electron saturation current for the case of argon is already shown in figure-\ref{Iesat}. Hence the number density of electrons can be obtained from the electron saturation current using the expression, $I_e=0.25en_e A_p V_{th}$. To benchmark the electron density obtained using LP, additionally a DC biased hairpin probe (HP) has been applied. The details about the DC biased HP technique can be found in our earlier work \cite{Pandey2020}. The objective of applying DC bias on the HP is to make the sheath around the cylindrical pins to zero, such that the HP measures the absolute value of electron density. Using this independent method, the electron density obtained is plotted for the case of argon plasma as shown in figure-\ref{LPHP}. As it can be seen that the electron density are found to be in very close agreement.

\begin{figure*}[hbp]
  \centering
     \includegraphics[width=0.5\textwidth]{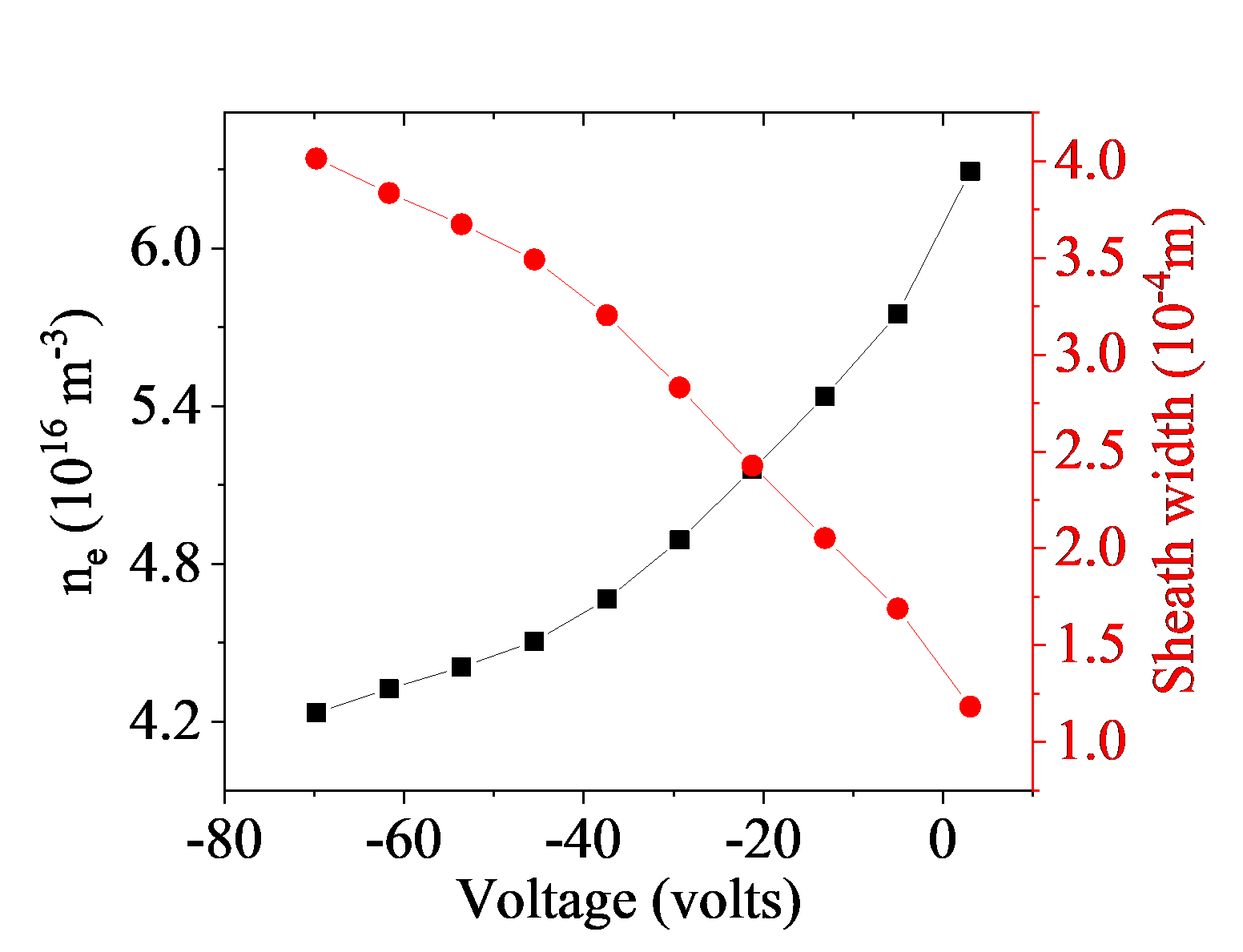}
    \caption{\label{sheathwidth} Plot of electron number density ($n_e$) and sheath width (s) versus the applied voltage at 130 W of power.}
  
\end{figure*}

The DC biased HP can also provide the information about the sheath width variation around the cylindrical wire surface as a function of the applied bias to the HP. The methodology to extract sheath using DC biased HP is described in our earlier work \cite{Pandey2020}. The sheath width and electron density obtained as a function of negative bias voltage on the HP for the case of argon is shown in figure-\ref{sheathwidth}. From the plots, it can be seen that the sheath width tends to vanish as the bias voltage approaches the plasma potential i.e $V_b  = V_p$; where $V_p$  = 5.3 Volts. 

With above information of sheath width variation around the cylindrical wire, the sheath width correction factor k can be found in equation-\ref{thr3}. The sheath width is also introduced in the positive ion saturation current, to eliminate the effect of increasing ion saturation current in figure-\ref{IVcorr}. From this the actual ion saturation current can be estimated corresponding to the floating potential as shown for the case of argon in figure-\ref{Iisat}.

\begin{figure*}[bbp]
  \centering
     \includegraphics[width=0.5\textwidth]{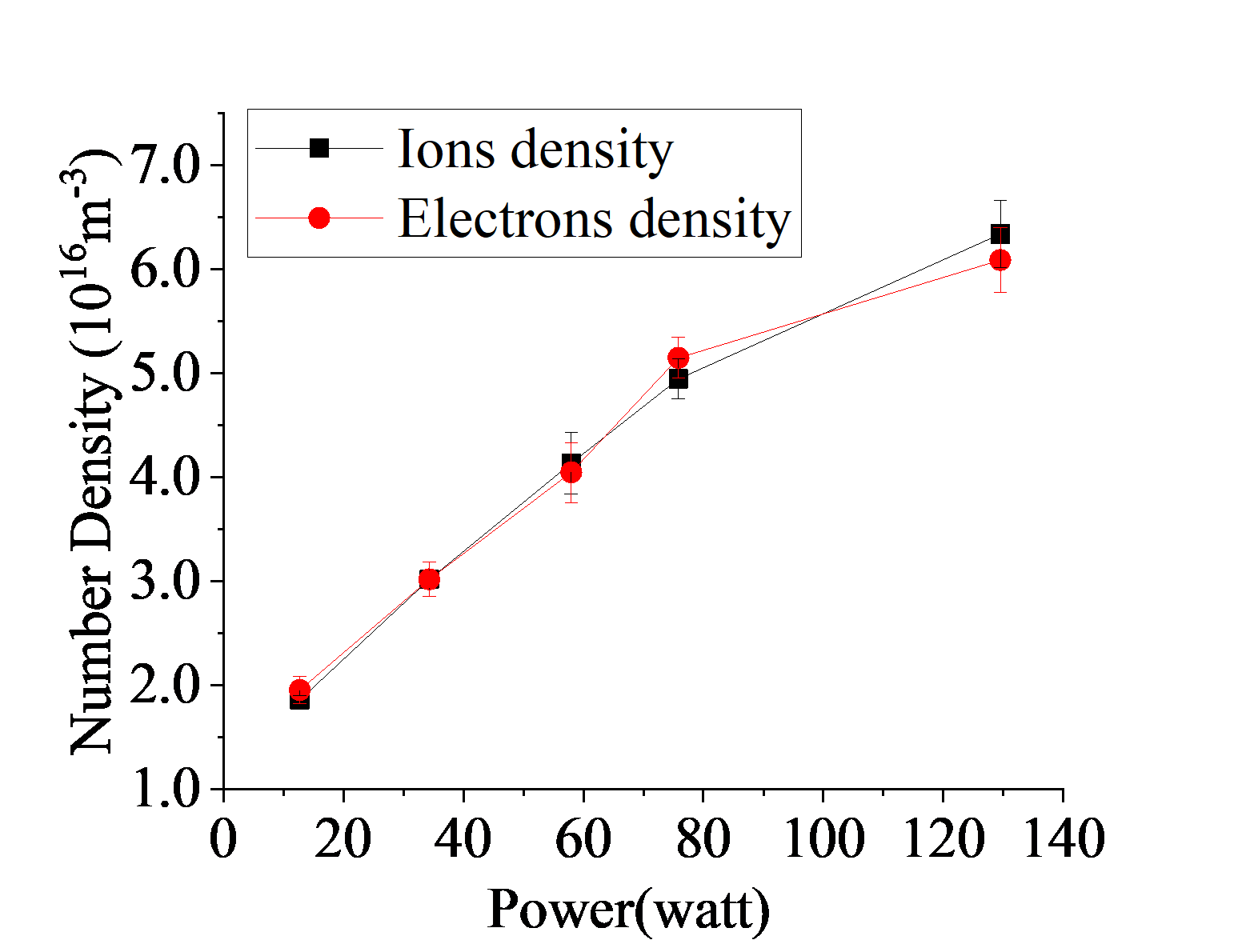}
    \caption{\label{nenp} Plot of electron and ion number density calculated using electron and ion saturation current respectively versus Discharge Power.}
  
\end{figure*}

\begin{figure*}[bbp]
  \centering
     \includegraphics[width=0.5\textwidth]{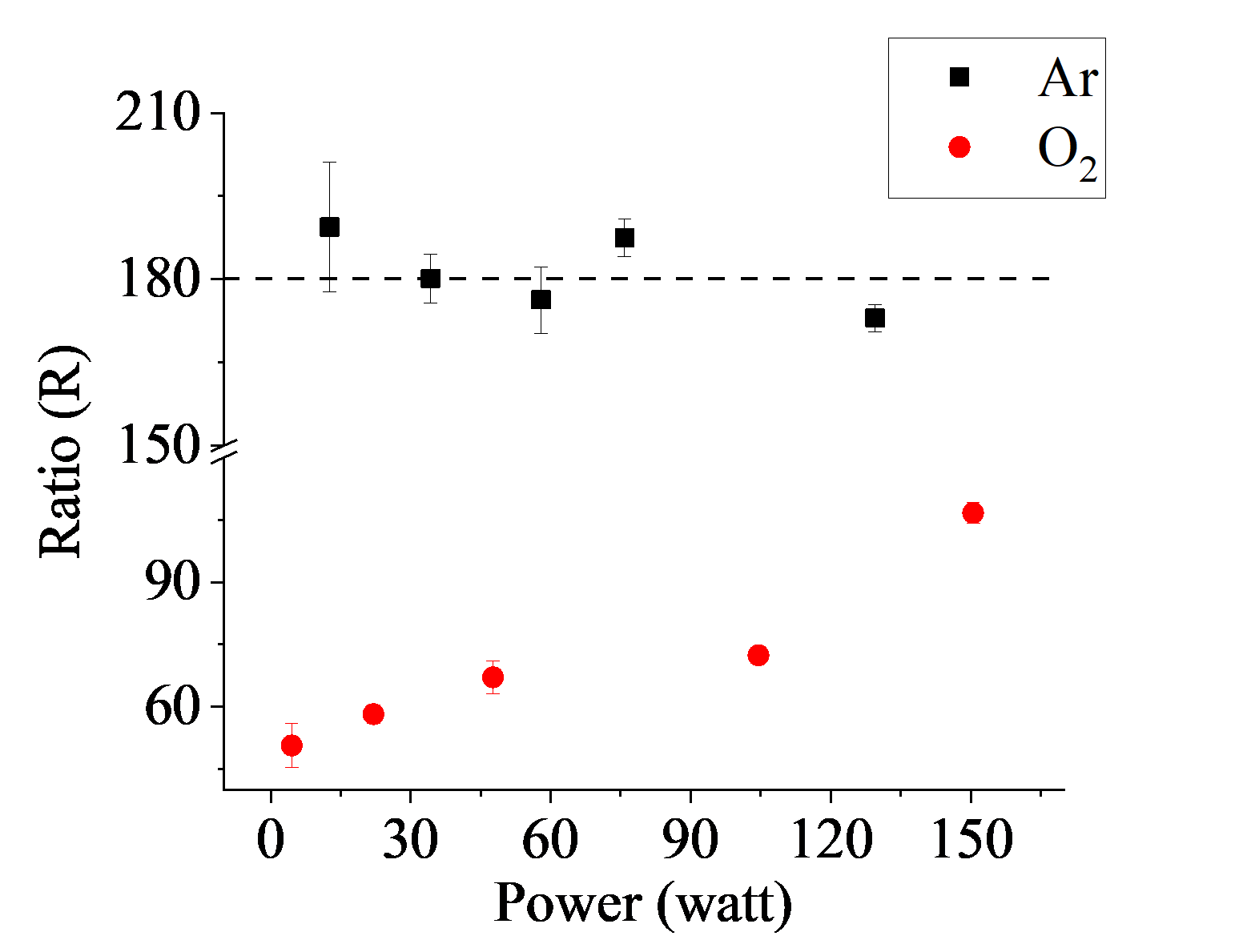}
    \caption{\label{Ratio} Plot of Ratio of electron saturation to ion saturation current versus Discharge Power for argon and oxygen plasma.}
  
\end{figure*}
By following the above procedure, the number density of electrons obtained from electron saturation current and positive ion density from the ion saturation currents using the cylindrical LP are plotted in figure-\ref{nenp}. It is quite apparent that for the case of electro-positive argon plasma having equal number of electrons and positive ions, both the densities should match each other.  Once the probe has been calibrated, then the electron and positive ion saturation currents are obtained for argon as well as oxygen plasma and their ratio are plotted in figure-\ref{Ratio}. As seen for the case of argon plasma, the values of $\frac{I_{es}}{I_{+s}}$ is close to the theoretically value 180, which remain to be fairly independent of the discharge power. Whereas in the case of electro-negative oxygen plasma, this ratio is significantly low during operation of the discharge at low pressures and tend to increase as the discharge power increases due to the contribution of the negative ions. 

\subsection{Models to determine $\alpha$ from Saturation Current Ratio}

In the literature, various authors have applied the Saturation Current Ratio method to determine the electronegativity parameter. However the underlying result strictly depend on the expression of electron and ion saturation currents which have been incorporated in the model to express the electronegativity parameter as function of the saturation current ratio. To exemplify this fact we present a few cases based on techniques adopted by different authors. 
  In the case of a highly electronegative plasma, ($\alpha \gg 1$), Sheridan et al. \cite{Sheridan1999} used the equation-\ref{thr5} to determine the presheath potential fall, which is then substituted in equation-\ref{thr4} to evaluate the Bohm velocity at the probe sheath edge. This gives the corresponding expression for the positive ion saturation current as follows, 

\begin{eqnarray}
I_{i,sat}=\frac{n_{p0}eA_{sheath}}{\gamma_v} \sqrt{\frac{eT_e}{M}}[(\frac{1}{2})^{\frac{1}{\gamma_v}}+\frac{\alpha}{2}] \label{expres1}
\end{eqnarray}

In the above expression, $A_{sheath}$   is the area of the cylindrical sheath around the probe surface. Since the sheath is itself a function of bulk plasma parameter as shown in figure-\ref{sheathpot}; therefore for simplicity, Bowes et al. \cite{Bowes2014} (method-1) assumed it to be equivalent to the probe area. This could be a valid approximation for a thin sheath in which $r_{probe}$ $\gg$ s.

Bowes et al. \cite{Bowes2014} have further introduced a comparative method to relate the electronegativity parameter in an electronegative oxygen plasma in terms of $\psi=\frac{\frac{I_{es}}{I_{ps}}_{O_2}}{\frac{I_{es}}{I_{ps}}_{Ar}} = \frac{R_{O_2}}{R_{Ar}}$   and $\gamma_v$, obtained by comparing the saturation current ratio with a known electro-positive argon plasma. This is expressed as follows

\begin{eqnarray}
\alpha=\frac{\sqrt{\gamma_v}}{\psi}-2^{\frac{\gamma_v-1}{\gamma_v}} \label{expres2}
\end{eqnarray}

However the probe sheath resistance to ground can vary differently for argon and oxygen plasma, hence absolute calibration using the above method cannot be fully assured.

To incorporate the probe sheath area in the ion saturation current, Bredin et al. \cite{Bredin2014} (method-2) used the Child Langmuir sheath model (CLSM) to calculate the sheath width as discussed in section-\ref{thrcorr}. In this model, the positive ion speed and density at the sheath edge must be supplied as the boundary condition. But the sheath parameters $n_s$ and $u_B$ are itself dependent on $\alpha$. To circumvent this dependency, Bredin et al. \cite{Bredin2014} introduced an iterative method which is briefly summarized as follows:

In this method, the bulk electron density is determined from the electron saturation current obtained at $V_p$; whereas $T_e$ from electron energy distribution function as expressed in equation-\ref{expres3} and equation-\ref{expres4}. 

\begin{eqnarray}
n_e=\frac{I_{e,sat}}{0.25eA_pv_{th,e}} \label{expres3}
\end{eqnarray}

\begin{eqnarray}
T_e=\frac{2}{3n_e}\int_{0}^{\infty}\varepsilon \sqrt{\varepsilon} f_e(\varepsilon) d\varepsilon \label{expres4}
\end{eqnarray}

To calculate the Bohm speed at the sheath edge, an initial guess value of $\alpha$ is supplied to equation-\ref{thr4} whereas the initial sheath width (s) = 0 is assumed to be zero. The positive ion density corresponding to a given value of  $I_{i,sat}$ is obtained as follows, $n_p=\frac{I_{i,sat}}{h_reA_su_B}$. This ion density $n_p$ and $u_B$ is then introduced in CLSM to evaluate the sheath width; whereas the quasi-neutrality condition, $n_p=n_e+n_v=(1+\alpha)n_e$ , is applied to get the value of $\alpha$. This sheath width and $\alpha$ is again introduced in equation-\ref{thr4} to evaluate the modified bohm speed and the positive ion density at the sheath edge. This process is iterated until a saturation value of $\alpha$ is reached.

Though this is an elegant method, however errors can result due to underestimation in electron density in the first place. Also, as the CLSM does not take into account the collisions and is sensitive to the boundary conditions, therefore estimated sheath width could be imprecise.

Another use of the saturation ratio method to determine $\alpha$ has been recently demonstrated by Pandey et al. \cite{Pandey2017} (method-3). In this method, the electron to positive ion saturation current ratio is expressed according to equation-\ref{thr1}; however in that model the probe sheath area is taken as the probe area to ignore the effect of sheath around the cylindrical wire. Doing so, equation-\ref{thr8} can be written as;

\begin{eqnarray}
(R^2 - \frac{\mu_{pv}}{2\pi 0.6^2}) \alpha^3  + [ 3R^2 -\frac{\mu_{pv}}{2\pi 0.6^2\gamma_v} -2\frac{\sqrt{\mu_{pv}\mu_{pr}\gamma_v}}{2\pi 0.6^2}] \alpha^2\nonumber\\ + [3R^2- \frac{2}{2\pi 0.6^2} \sqrt{\frac{\mu_{pv}\mu_pe}{\gamma_v}} - \frac{\gamma_v\mu_{pe}}{2\pi 0.6^2}] \alpha + R^2 - \frac{\mu_{pe}}{2\pi 0.6^2}=0 \label{expres6}
\end{eqnarray}

The above equation can be numerically solved for a given value of R obtained from LP. However there are two major limitation in this model. First the sheath edge density was assumed to be related to the bulk plasma density by a factor 0.6. This is only applicable for the case of electropositive argon plasma and is a function of $\alpha$ for the electronegative oxygen plasma. Furthermore in order to simplify the equation, the negative ion density around the probe sheath, i.e. $\alpha_s$ had been approximated to be same as $\alpha$. This approximation may be valid for the case of highly electronegative plasmas, but the pre-sheath potential must also be accordingly found using the figures-\ref{etavsalpha} and figure-\ref{fluxvsalpha} as described in section-\ref{thrcorr}.

With the aforementioned issues, a refined analytical expression to determine $\alpha$ from the electron and ion saturation current ratio has been formulated. This is presented in section-\ref{thr}. The model takes in to account the potential drop across the pre-sheath which is related to the bulk electro-negativity parameter inside the plasma. The methodology behind using the improved equation to find the negative ion density parameter is as follows;

In the first step, the saturation current ratio R is accurately determined as presented above. To find the true value of ion saturation current, the probe sheath area correction factor k is determined from HP. The parameter $\gamma_v$ which is a ratio of electron to negative ion temperature is calculated by assuming a constant negative ion temperature value which is typically in the range of 0.05 eV as found in the literature \cite{Sirse2015,Pandey2020}; and the electron temperature is determined from LP. Then equations-\ref{thr4} and \ref{thr5} are combined to obtain a relation between $\alpha$ and $\eta_s$. For a particular value of $\gamma_v$ = 40, this has been plotted in figure-\ref{etavsalpha}. Furthermore, substituting $\eta_s$ in equation-\ref{thr3}, the positive ion flux at the sheath boundary can be obtained as a function of $\alpha$, which is plotted in figure-\ref{fluxvsalpha}. Since the potential at the pre-sheath is a triple value solution, therefore one has to take the solution corresponding to the maximum value of positive ion flux for a given value of $\alpha$ as shown from figure-\ref{fluxvsalpha}.

In the underlying equations, the parameter $\eta_s$ and $\alpha$ is inter-dependent. Therefore a guess value of $\alpha$ is initially assumed to determine the pre-sheath potential drop $\eta_s$. It is then used to calculate a revised value of $\alpha$ using equating-\ref{thr8}. The revised $\alpha$  is further used to determine the pre-sheath potential and this process is repeated until a converging value of $\alpha$ is obtained. To verify this iterative method, a resonance hairpin probe is applied to determine the electronegativity parameter $\alpha$ by comparing the electron density with the sheath compensated ion saturation current as articulated in reference.

\begin{figure*}[!hbp]
  \centering
     \includegraphics[width=0.5\textwidth]{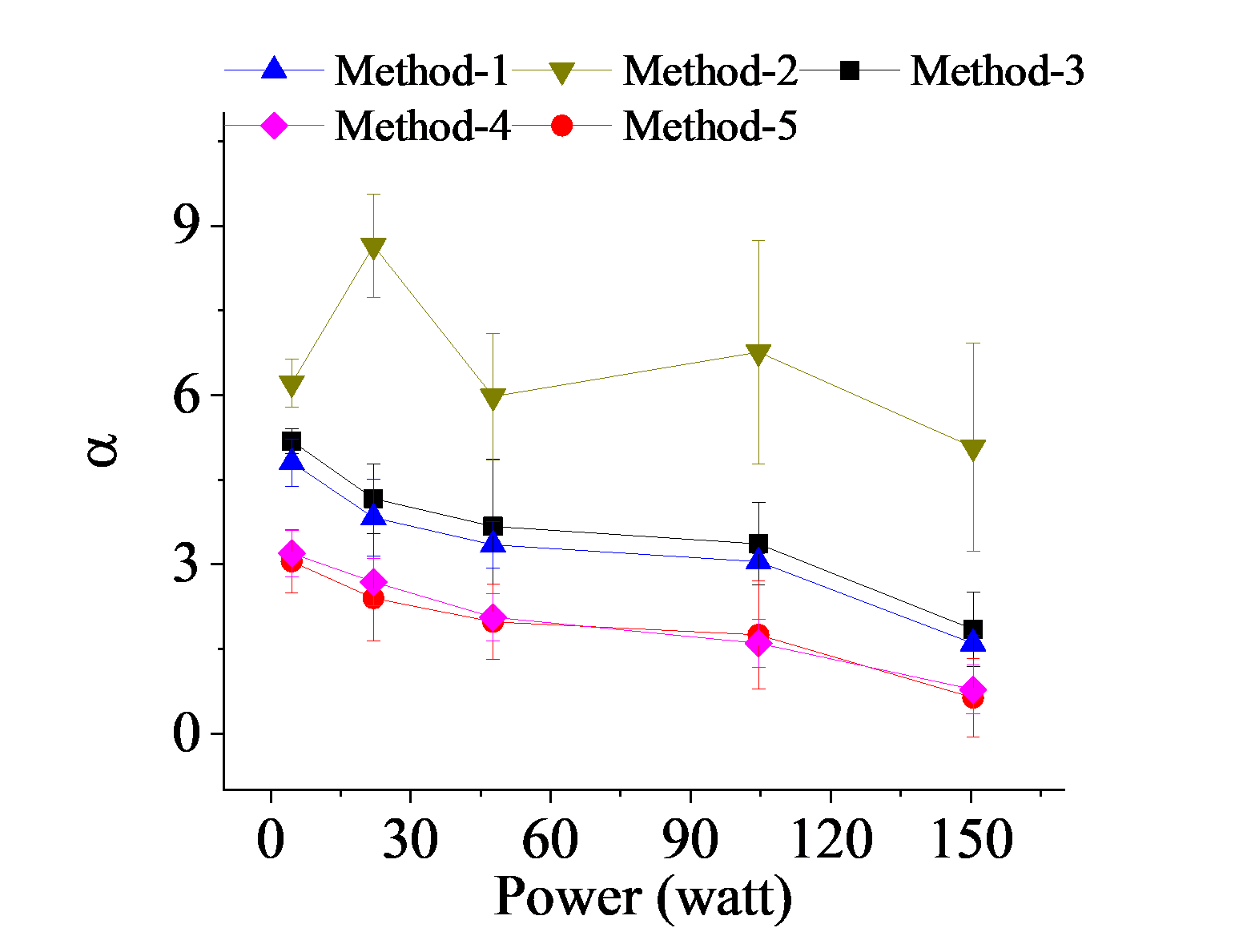}
    \caption{\label{result} Plot of alpha (ratio of negative ion density to electron density) versus Discharge power versus Discharge Power using varies techniques. Method-1 = Bowes et al., method-2 = Bredin et. al., method-3 = Avnish et al., method-4 = present model and method-5 = biased HP.}
  
\end{figure*}

In figure-\ref{result}, the electronegativity parameter $\alpha$ obtained using different methods are plotted corresponding to a range of discharge power. The plots reveal a similar trend that the negative ion density decrease as the discharge power increases. This is possibly due to increase in electron impact detachment \cite{Pandey2017}. It can be seen that the refined ratio technique (method-4) and the $\alpha$ determined using the HP (method-5) are very well is agreement; however all the previous methods shows an overestimated values of $\alpha$. The methodology by Bowes et al. \cite{Bowes2014} and Pandey et al. \cite{Pandey2017} seems to be in agreement as they are based on similar approximation. A maximum uncertainty in the value of $\alpha$ is seen in the case of Bredin et al. \cite{Bredin2014}. In Bredin’s model, a CLSM has been applied for the calculation of sheath, however the model is dependent on the boundary conditions at the sheath edge which is subject to errors as well as the model does not account for the collisions in the presheath. Different authors had shown that the ion saturation current can be reduced to as much as half even for weakly collisional plasma \cite{Bryant2001,Sudit1994,SHIH1971}; therefore, applying CLSM for the calculation of the sheath width may lead to erroneous results.

\section{Summary:}\label{Sum}
As the paper demonstrates, determining negative ion parameter using LP is subjected to significant deviation due to various approximations which has gone in the underlying models adopted by different authors. This is because the saturation currents collected by the probe is an implicit function of the sheath edge density and the positive ion speed, which itself are dependent on the bulk negative ion parameters to be found from the experiment. In fact there is no direct means to estimate the sheath width using LP but to use an analytical sheath model like CL, however the model is very much sensitive to the boundary conditions that require the information of these plasma parameters which one is interested to find. In this paper a hairpin probe has been used to estimate the sheath width variation as a function of negative bias; and with this information it is possible to correct the positive ion saturation current measured by the LP, by incorporating the effective probe surface area with negative bias with respect to $V_p$. The potential drop across the pre-sheath dictate the positive ion speed at the sheath boundary, which directly impacts the positive ion flux collected by the cylindrical probe. As highlighted, a number of authors have assumed the pre-sheath potential to be equivalent to the case of an electro-positive plasma, bult clearly such approximations has a tremendous impact on the obtained values of  $\alpha$. On the other hand, estimation of electron saturation current from a cylindrical LP has a limitation, mainly due to uncertainty in locating the plasma potential in the I(V) characteristics. In certain cases the electron current tends to saturate at a much lower values due to ground sheath resistance or during application or external magnetic field. To avoid these limitations, some authors have prescribed calibration of the saturation current ratio with the help of a known electro-positive plasma like argon. However this calibration is valid only if the nature of the sheath resistance at the reference electrode remains unchanged. In-spite of all the shortfalls associated with LP to determine $\alpha$, but so far the underlying limitations behind them had not been addressed comprehensively. This paper is by far the first attempt to highlight the various anomalies associated with the existing methods to determine electronegativity parameter using the saturation current ratio method; and suggest fundamental steps that must be taken in to account for applying this technique for the determination of negative ions in low pressure electronegative discharges.

\section*{References}

\bibliographystyle{unsrt}
\bibliography{LPpaper}

\end{document}